\documentclass{article}
\usepackage[T1]{fontenc}
\usepackage[utf8]{inputenc}
\usepackage{prettyref}
\usepackage{float}
\usepackage{amsmath}
\usepackage{graphicx}
\usepackage[authoryear]{natbib}

\makeatletter

\let\SF@@footnote\footnote
\def\footnote{\ifx\protect\@typeset@protect
    \expandafter\SF@@footnote
  \else
    \expandafter\SF@gobble@opt
  \fi
}
\expandafter\def\csname SF@gobble@opt \endcsname{\@ifnextchar[
  \SF@gobble@twobracket
  \@gobble
}
\edef\SF@gobble@opt{\noexpand\protect
  \expandafter\noexpand\csname SF@gobble@opt \endcsname}
\def\SF@gobble@twobracket[#1]#2{}

\usepackage{prettyref}
\newrefformat{eq}{eqn\,(\ref{#1})}
\newrefformat{sec}{Section\,\ref{#1}}
\newrefformat{cha}{Chapter\,\ref{#1}}
\newrefformat{tab}{Table\,\ref{#1}}
\newrefformat{fig}{Figure\,\ref{#1}}
\newrefformat{lem}{Lemma\,\ref{#1}}
\newrefformat{thm}{Theorem\,\ref{#1}}
\newrefformat{axm}{Axiom\,\ref{#1}}
\newrefformat{app}{Appendix\,\ref{#1}}
\newrefformat{exe}{Exercise\,\ref{#1}}
\newrefformat{cor}{Corollary\,\ref{#1}}
\newrefformat{pro}{Property\,\ref{#1}}
\newrefformat{def}{Definition\,\ref{#1}}
\newrefformat{subsec}{Section\,\ref{#1}}
\usepackage{amstext}
\usepackage{amsmath}
\usepackage{bm}
\numberwithin{equation}{section}
\numberwithin{figure}{section}
\numberwithin{table}{section}
\newcommand{\br}[1]{\mathbf{#1}} 
\newcommand{\ms}[1]{\mathsf{#1}} 
\newcommand{\bb}[1]{\mathbb{#1}} 
\newcommand{\ca}[1]{\mathcal{#1}} 
\newcommand{\mat}[1]{\ms{#1}} 
\newcommand{\op}[1]{\ca{#1}} 
\usepackage{pifont}
\usepackage{txfonts}
\usepackage{charter}
\date{}

\makeatother

\begin{document}

\title{Validity of the Einstein Hole Argument\thanks{Published in \emph{Studies
in the History and Philosophy of Modern Physics} (2019) \textbf{68}: 62-70. }}

\author{{\normalsize{}Oliver Davis Johns}\\
{\normalsize{}San Francisco State University, Physics and Astronomy
Department}\\
{\normalsize{}1600 Holloway Avenue, San Francisco, CA 94133, USA}\\
{\normalsize{}Email: ojohns@metacosmos.org}\\
{\normalsize{}Web: http://www.metacosmos.org}}

\maketitle
\begin{abstract}
Arguing from his \textquotedbl{}hole\textquotedbl{} thought experiment,
Einstein became convinced that, in cases in which the energy-momentum-tensor
source vanishes in a spacetime hole, a solution to his general relativistic
field equation cannot be uniquely determined by that source. After
reviewing the definition of active diffeomorphisms, this paper uses
them to outline a mathematical proof of Einstein's result. The relativistic
field equation is shown to have multiple solutions, just as Einstein
thought. But these multiple solutions can be distinguished by the
different physical meaning that each metric solution attaches to the
local coordinates used to write it. Thus the hole argument, while
formally correct, does not prohibit the subsequent rejection of spurious
solutions and the selection of a physically unique metric. This conclusion
is illustrated using the Schwarzschild metric. It is suggested that
the Einstein hole argument therefore cannot be used to argue against
substantivalism.
\end{abstract}
\rule{0.4\textwidth}{0.2pt}
\begin{enumerate}
\item \emph{Introduction}
\item \emph{Preliminary Remarks}
\item \emph{Active and Passive Diffeomorphisms}
\item \emph{Einstein's Hole Argument in General Relativity}
\item \emph{Physical Meaning of Einstein's Multiple Metrics}
\item \emph{The Schwarzschild Example}
\item \emph{The Hole Argument with the Schwarzschild Solution}
\item \emph{Conclusion}
\end{enumerate}
~

\section{Introduction}

\label{sec:intro}Einstein's \textquotedbl{}hole\textquotedbl{} thought
experiment convinced him that specification of the energy-momentum-tensor
source would not determine a unique solution to his general relativistic
field equation.\footnote{See Chapter 5 of \citet{torretti} and Chapter 5 of \citet{stachel-btoz}
for the history of Einstein's quest for the equations of general relativity.} 

Einstein's own description of the argument was brief and lacking in
detail. He first refers to the required transformations as what translates
as  \emph{coordinate} transformations, and later as \emph{point} transformations.\footnote{See Section 5.6 of \citet{torretti}.}
\citet{Stachel-active} has interpreted this latter phrase as referring
to what he calls \textquotedbl{}active diffeomorphisms.\textquotedbl{}\footnote{Einstein's term would be \textquotedbl{}point diffeomorphism.\textquotedbl{}
I use the terms \textquotedbl{}active diffeomorphism\textquotedbl{}
and \textquotedbl{}point diffeomorphism\textquotedbl{} as exact synonyms. } In an attempt to avoid misunderstandings about notation and definitions,
\prettyref{sec:prelim} makes some preliminary remarks and \prettyref{sec:actpas}
uses basic differential geometry to define Stachel's term \textquotedbl{}active
diffeomorphism\textquotedbl{} and its companion term \textquotedbl{}passive
diffeomorphism.\textquotedbl{} 

Einstein posited a specific experimental situation in which a \textquotedbl{}hole\textquotedbl{}
region $H$ in spacetime is devoid of energy-momentum-tensor sources
($T_{\mu\nu}(x)\equiv0$ for $x\in H$), with this hole surrounded
by a source region $S$ in which the energy-momentum tensor could
be nonzero.\footnote{\citet{EinsteinGrossmann}. See also paraphrase by \citet{torretti},
p. 163.} He argued that an active diffeomorphism that acted as the identity
in the $S$ region, but was not an identity in the hole, would modify
the metric field in the hole without modifying any of the sources,
either inside or outside the hole. He concluded that the energy-momentum
sources cannot determine the metric field in the hole uniquely. \prettyref{sec:hole-in-gr}
outlines a proof of Einstein's conclusion.

But the existence of a mathematical proof that Einstein's field equation
has multiple solutions leads to the question of the physical meaning
of these multiple solutions.\footnote{By \textquotedbl{}physical meaning\textquotedbl{} (sometimes shortened
to just \textquotedbl{}meaning\textquotedbl{}) of a solution I mean
a set of defined relations between the local coordinates used to write
it and something like length or relativistic interval, such as is
defined by a Riemannian metric. } This issue is addressed in \prettyref{sec:phys}, which discusses
the difficulties introduced into differential geometry by Einstein's
disruptive idea of a Riemannian metric that is not known until after
a differential equation for it is solved. Before the field equation
is solved, since there is not yet a defined metric, the local coordinates
are just $m$-tuples of real numbers that have no definite relation
to anything physical like relativistic interval. After the field equation
is solved, each of the multiple solutions produced by the hole argument
is then a distinct metric that attaches its own distinct physical
meaning to the local coordinates that were used to write it. It may
thus be possible to select among the multiplicity of mathematical
solutions of \prettyref{sec:hole-in-gr} a unique one that assigns
to its local coordinates the physical meaning needed to model the
symmetries of the experimental situation under study, rejecting the
other metric solutions as spurious.  Thus the hole argument in \prettyref{sec:hole-in-gr}
fails to prove that Einstein's field equation must necessarily have
multiple non-spurious solutions.

Sections \ref{sec:schwarz} and \ref{sec:schole} illustrate these
ideas using the Schwarzschild solution for a spherically symmetric
source mass. In this case, a unique solution is found, thus providing
a counterexample to the proposition that the Einstein field can never
have a unique solution.

\section{Preliminary Remarks}

\label{sec:prelim}A few preliminary remarks may be helpful. First,
in discussing the uniqueness of solutions to generally covariant differential
equations, it is necessary to remember that any solution must be expressed
in some system of local coordinates. A solution written in one coordinate
system can, by a diffeomorphic change of local coordinates (passive
diffeomorphism as defined in \prettyref{subsec:passive}), always
be transformed into the same solution expressed in some other coordinate
system. (Think of converting from Cartesian to spherical polar coordinates
in Euclidean three-space.) But the existence of these two expressions
in the two coordinate systems is not what is meant when one speaks
of non-uniqueness of solution. These are not \emph{different} solutions,
but only the \emph{same} solution expressed in two different systems
of local coordinates. To say that a generally covariant differential
equation has a second solution and therefore is non-unique means a
second solution that is different from the original one when \emph{both}
solutions are expressed in the \emph{same} local coordinate system.
This is the sense in which Einstein used the term \textquotedbl{}unique,\textquotedbl{}
and also the sense in which it is used in this paper.

Second, it is necessary to realize that there are at least two distinct
and non-equivalent definitions of the hole argument extant in the
literature. The first is that due to Einstein outlined above. There
is no evidence that Einstein ever intended his hole argument to apply
to generally covariant differential equations other than his own general
relativistic field equation. Also, as will be shown in \prettyref{sec:hole-in-gr},
Einstein's version depends essentially on his assumption of a particular
experimental situation in which the energy momentum tensor term $T_{\mu\nu}(x)$
in his differential equation vanishes identically in the region he
calls the \textquotedbl{}hole.\textquotedbl{} 

On the other hand, the revision of the hole argument by Earman and
Norton\footnote{\citet{earman-norton,norton-encyc}} is asserted by
them to include \textquotedbl{}...Newtonian spacetime theories with
all, one, or none of gravitation and electrodynamics; and special
and general relativity, with and without electrodynamics.\textquotedbl{}\footnote{However, the generally covariant Poisson equation for the electrostatic
potential in three dimensions, when applied to a spherically symmetric
source with the generally covariant boundary condition that the potential
vanish at infinity, is well known to have a unique solution, thus
providing a counterexample to Earman and Norton’s assertion that their
hole argument applies also to such differential equations.} Also their presentation of the hole argument does not require that
a source term must vanish in the hole region. They assert that Einstein's
presentation is, only \textquotedbl{}$\ldots$a specialized form$\ldots$\textquotedbl{}
of their generalized hole argument.\footnote{Earman and Norton, op.cit., p. 523. The Earman-Norton version explores
the consequences of a Leibnizian interpretation of active diffeomorphisms.
It makes no direct reference to the details of the Einstein field
equation, which details are the main focus of the Einstein version
studied in the present paper.} 

This existence of two distinct hole arguments has confused the subject,
with some refutations of what their authors take to be\emph{ the }hole
argument apparently applying only to the Earman-Norton version.\footnote{See for example the recent articles: \citet{weatherall,schulman-homotopy}}
This paper will not derive or defend the Earman-Norton version.

The third preliminary remark concerns style. It has become common
to discuss the hole argument in abstract mathematical language.\footnote{For example, the use of category theory in \citet{IftimeStachel}.}
But the subtlety of Einstein's argument is revealed only when one
uses coordinates to study it. Fortunately, although invariant language
is the norm today, arguments using coordinates are not therefore invalid.
They may seem crude, but they are still true.

There is an analogy here to computer programming languages. High-level
languages such as Python or C++ are elegant and succinct, but every
programmer knows that there are some problems that require low-level
machine assembly language to solve. In this paper, I discuss the Einstein
hole argument using high-level invariant language—and assembly language
when required.

\section{Active and Passive Diffeomorphisms}

\label{sec:actpas}This section outlines the definition of the term
\textquotedbl{}active diffeomorphisms\textquotedbl{} and gives a method
for generating them.

\begin{figure}[h]
\centerline{\includegraphics[scale=0.85]{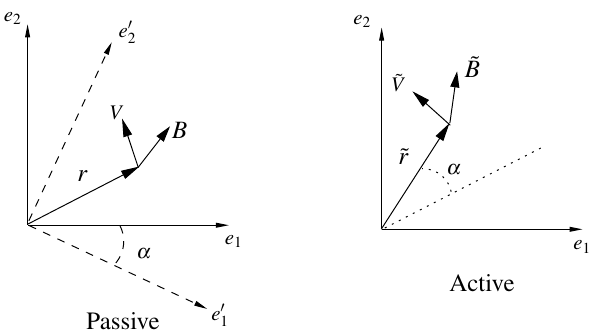}}

\caption{\label{fig:Act-Pass-fig}Active and passive transformations. The passive
one rotates the coordinate system (observer) but leaves the vectors
(physical world) unchanged. The active transformation rotates the
vectors but leaves the coordinate system unchanged.}
\end{figure}

The distinction between active and passive diffeomorphisms is borrowed
from the transformation theory of classical vector calculus.\footnote{For example, in Chapter 8 of \citet{oj} active transformations are
used initially and passive transformations are introduced in Section
8.30.} Suppose that a three-dimensional Euclidean coordinate system containing
a velocity $\br V$ and another field $\br B$ at point $\br r$ is
rotated by angle $\alpha$ about the $z$-axis as shown on the left
side of \prettyref{fig:Act-Pass-fig}. Suppose that before rotation,
the components of the vectors are $\br r:(x^{1},x^{2},x^{3})$, $\br V:(V^{1},V^{2},V^{3}{}_{)}$,
and $\br B:(B^{1},B^{2},B^{3})$. After the rotation the vectors are
unchanged, but their components become $\br r:(x'^{1},x'^{2},x'^{3})$,
$\br V:(V'^{1},V'^{2},V'^{3})$, and $\br B:(B'^{1},B'^{2},B'^{3})$
where
\begin{align}
x'^{1}=x{}^{1}\cos\alpha-x{}^{2}\sin\alpha\quad\quad & x'^{2}=x{}^{1}\sin\alpha+x{}^{2}\cos\alpha\quad\quad x'^{3}=x^{3}\label{eq:a1}\\
V'^{1}=V{}^{1}\cos\alpha-V^{2}\sin\alpha\quad\quad & V'^{2}=V{}^{1}\sin\alpha+V{}^{2}\cos\alpha\quad\quad V'^{3}=V^{3}\label{eq:a2}
\end{align}
with similar expressions for the components of $\br B$. The observer,
here represented by the coordinate system, rotates by angle $\alpha$
but the physical world being observed, here represented by the vectors,
does not rotate. This is called a passive transformation since the
world is not changed, just the view of the observer.

An active transformation rotates the physical world by angle $\alpha$
about the $z$-axis while keeping the observer fixed, as shown on
the right side of \prettyref{fig:Act-Pass-fig}. The observer's coordinate
system is not changed, but the vectors are changed to new vectors
$\br{\tilde r}$, $\br{\tilde V}$, and $\br{\tilde B}$, with components
(expressed in the unchanged original coordinate system)
\begin{align}
\tilde{x}^{1}=x^{1}\cos\alpha-x^{2}\sin\alpha\quad\quad & \tilde{x}^{2}=x^{1}\sin\alpha+x^{2}\cos\alpha\quad\quad\tilde{x}^{3}=x^{3}\label{eq:a3}\\
\tilde{V}^{1}=V^{1}\cos\alpha-V^{2}\sin\alpha\quad\quad & \tilde{V}^{2}=V^{1}\sin\alpha+V^{2}\cos\alpha\quad\quad\tilde{V}^{3}=V^{3}\label{eq:a4}
\end{align}
 with similar expressions for $\br{\tilde B}$. This is called an
active transformation since the observed world is changed but the
observer is kept fixed.

\subsection{Passive Diffeomorphism}

\label{subsec:passive}The obvious differential geometric analog of
the classical passive transformation of vector components in \prettyref{eq:a1}
is the diffeomorphic change of local coordinates on a smooth manifold.\footnote{See \citet{lee-smooth,lee-top}, texts I take to be the canonical
references for modern, invariant differential geometry.} 

Let a manifold $\op M$ of dimension $m$ have two overlapping charts
of local coordinates $(\psi,U)$ and $(\psi',U')$ where $U$ and
$U'$ are open sets in $\op M$ with $U\cap U'\neq0$, and $\psi$,
$\psi'$ are homeomorphisms from $U$, $U'$ to local coordinates
$x=(x^{1},\ldots,x^{m})$ and $x'=(x'^{1},\ldots,x'^{m})$, respectively,
in $\bb R^{m}$. If the function $x'=\gamma(x)$, where $\gamma=\psi'\circ\psi^{-1}$,
and its inverse $x=\gamma^{-1}(x')$ are both continuously differentiable
to arbitrary order for all such overlapping open sets $U$,$U'$,
the manifold $\op M$ is a smooth manifold and $\gamma:x\rightarrow x'$
is a diffeomorphic change of local coordinates. A point $p\in\op M$
is represented either by local coordinates $x=\psi(p)$ or $x'=\psi'(p)$.
Such \textquotedbl{}diffeomorphic changes of local coordinates\textquotedbl{}
are referred to in this paper as \textquotedbl{}passive diffeomorphisms.\textquotedbl{}\footnote{The term \textquotedbl{}gauge transformations of differential geometry\textquotedbl{}
is also sometimes used in the literature.}

Smooth functions $f:\op M\rightarrow\bb R$ mapping points $p$ on
the smooth manifold to real numbers $f(p)$ are represented in unprimed
and primed local coordinates by $F=f\circ\psi^{-1}$ and $F'=f\circ\psi'^{-1}$
so that\footnote{In the literature, function $F(x)$ is often written $f(x)$. One
is supposed to read from the variable, $x$ rather than $p$, that
$F$ is intended. The condition that $f$ be a smooth function is
that local coordinate function $F(x)$ must be continuously differentiable
to arbitrary order.} 
\begin{equation}
F(x)=f(p)=F'(x')\label{eq:a5}
\end{equation}
One must distinguish between manifold objects\footnote{The terms \textquotedbl{}manifold object\textquotedbl{} and \textquotedbl{}invariant
object\textquotedbl{} are used as synonyms in this paper. Manifold
objects like $f(p)$ are invariant under changes of local coordinates.} like $f(p)$ and local coordinate objects like $F(x)$. In physical
theories, manifold objects can be taken as real while local coordinate
objects only \emph{represent} the underlying manifold ones in various
local coordinate systems. 

A tangent vector field $\br V(p)$ is a manifold object in the tangent
bundle of $\op M$, a member of the tangent space over manifold point
$p$. Its action is represented in operator notation; it maps smooth
functions $f(p)$ to invariant real numbers denoted as $\br V(p)f(p)$.
It is represented in the unprimed and primed charts by
\begin{equation}
V(x)=\sum_{j=1}^{m}V^{j}(x)E_{j}\quad\quad\quad\quad V'(x')=\sum_{i=1}^{m}V'{}^{i}(x')E'_{i}\label{eq:a6}
\end{equation}
respectively, where $E_{j}=\partial/\partial x^{j}$ and $E'_{i}=\partial/\partial x'^{i}$
are local coordinate representations of basis vectors in the two charts.
Then
\begin{equation}
V(x)F(x)=\br V(p)f(p)=V'(x')F'(x')\label{eq:a7}
\end{equation}
and the components are related by the rule
\begin{equation}
V'^{i}(x')=\sum_{j=1}^{m}\dfrac{\partial x'^{i}}{\partial x^{j}}V^{j}(x)\label{eq:a8}
\end{equation}

A covariant tensor field of rank $k$ is a manifold object $\br g(p)$
that maps an ordered set of tangent vector fields to an invariant
real number denoted
\begin{equation}
\br g(p)\bigl\{\br V_{1}(p),\ldots,\br V_{k}(p)\bigr\}\label{eq:a9}
\end{equation}
If the manifold $\op M$ is Riemannian\footnote{In this paper, Riemannian always is intended to include Semi-Riemannian.}
with second rank, covariant metric tensor field $\br g(p)$, denoted
$(\op M,\br g)$, the invariant inner product of two tangent vector
fields is defined as 
\begin{equation}
\Bigl\langle\br V(p),\br W(p)\Bigr\rangle=\br g(p)\bigl\{\br V(p),\br W(p)\bigr\}\label{eq:aa10}
\end{equation}

The metric tensor field is represented in the unprimed and primed
charts by components $g_{ij}(x)$ and $g_{kl}'(x')$, respectively.
The inner product is then
\begin{equation}
\sum_{i,j=1}^{m}g_{ij}(x)V^{i}(x)W^{j}(x)=\Bigl\langle\br V(p),\br W(p)\Bigr\rangle=\sum_{k,l=1}^{m}g{}_{kl}'(x')V'^{k}(x')W'^{l}(x')\label{eq:aa11}
\end{equation}
and the local components of $\br g(p)$ transform as 
\begin{equation}
g_{ij}(x)=\sum_{k,l=1}^{m}g_{kl}'(x')\dfrac{\partial x'{}^{k}}{\partial x{}^{i}}\dfrac{\partial x'{}^{l}}{\partial x{}^{j}}\label{eq:aa12}
\end{equation}

\subsection{Active Diffeomorphism}

\label{subsec:act}In \prettyref{subsec:passive}, the differential
geometric analog of classical vector passive transformations was easily
available; one simply identified \textquotedbl{}passive diffeomorphisms\textquotedbl{}
with universally accepted definition of \textquotedbl{}diffeomorphic
change of local coordinates.\textquotedbl{} But the differential geometric
analog of the active transformation of classical vectors in \prettyref{eq:a3},
to be called an \textquotedbl{}active diffeomorphism\textquotedbl{}
here, is less well established and requires some definition. Some
texts on differential geometry for the general relativity community,
e.g., \citet{carroll,wald}, discuss active diffeomorphisms peripherally,
but other standard references on differential geometry for the pure
mathematics and high-energy physics communities, e.g., \citet{frankel,lee-riemann,lee-smooth,oneill,Taubes},
 do not even contain the phrase. However, they do contain a construction
that can be tailored to our purposes, the differentiable mapping $\phi:\op M\rightarrow\op N$
between two manifolds $\op M$ and $\op N$ of dimension $m$ and
$n$, respectively, where in general the dimensions are different,
$m\ne n$, and the mapping need not be a homeomorphism (a continuous
mapping with a continuous inverse).\footnote{See Chapters 2 and 3 of \citet{lee-smooth}.} 

Here we consider the restricted case in which $m=n$ and $\phi$ is
an active diffeomorphism. Thus, if $(U,\psi)$ and $(\tilde{U},\tilde{\psi})$
are charts of local coordinates $x$, $\tilde{x}$ on $\op M$ and
$\op N$, respectively, we assume that both $\tilde{x}=\theta(x)$,
where $\theta=\tilde{\psi}\circ\phi\circ\psi^{-1}$, and its inverse
$x=\theta^{-1}(\tilde{x})$ exist and are continuously differentiable
to arbitrary order.\footnote{It will be assumed uncritically here that the domains of the homeomorphisms
$\psi$, $\psi'$, and $\tilde{\psi}$ which define the local coordinates
comprise the whole of their respective manifolds. If multiple domains
are required in a particular case, it is assumed that they can be
patched together by standard techniques.}

Let $p$ and $\tilde{p}=\phi(p)$ be points in $\op M$ and $\op N$,
respectively, and let $\tilde{f}(\tilde{p})$ be a smooth function
$\tilde{f}:\op N\rightarrow\bb R$. Then there is a smooth function
$f(p)$ with $f:\op M\rightarrow\bb R$ defined by $f=\tilde{f}\circ\phi$.
This $f$ is called the \emph{pull-back} of $\tilde{f}$ and is denoted
$f=\phi^{*}\tilde{f}$. Since $\phi$ is assumed here to have an inverse,
the $\tilde{f}$ can also be written as what is called the \emph{push-forward}
of $f$, denoted $\tilde{f}=\phi_{*}f$. No matter how denoted, the
relation is 
\begin{equation}
\tilde{f}(\tilde{p})=f(p)\label{eq:aa13}
\end{equation}

This relation can also be written in local coordinates. If $\tilde{F}(\tilde{x})$
is a smooth function defined in terms of local coordinates on $\op N$,
then there is a smooth function $F(x)$, where $F=\tilde{F}\circ\theta$,
similarly defined on $\op M$. This $F$ is called a pull-back of
$\tilde{F}$ and is denoted $F=\phi^{*}\tilde{F}$. Since we are assuming
$\phi$ and hence $\theta$ to have an inverse, we can also refer
to $\tilde{F}$ as what is called a push-forward of $F$ denoted $\tilde{F}=\phi_{*}F$.
In either case, the relation is 
\begin{equation}
F(x)=\tilde{F}(\tilde{x})\label{eq:aa14}
\end{equation}
 which shows that $\tilde{F}(\tilde{x})$ at a point $\tilde{x}$
has the same value as function $F(x)$ has at point $x$. 

In general, since $\phi$ and $\theta$ are assumed to be diffeomorphic
here, and all transformations therefore possess inverses, both pull-back
and push-forward of functions, tangent vectors, and general tensor
fields are well defined. 

Tangent vector fields can also be pulled back or pushed forward. Let
$\br V(p)$ and $\tilde{\br V}(\tilde{p})$ be manifold objects on
$\op M$ and $\op N$, respectively. Then $\br V=\phi^{*}\tilde{\br V}$,
or equivalently $\tilde{\br V}=\phi_{*}\br V$, is defined by
\begin{equation}
\br V(p)f(p)=\tilde{\br V}(\tilde{p})\tilde{f}(\tilde{p})\label{eq:aa15}
\end{equation}
In terms of local coordinates with $\tilde{x}=\theta(x)$, this is
\begin{equation}
V(x)F(x)=\tilde{V}(\tilde{x})\tilde{F}(\tilde{x})\label{eq:aa16}
\end{equation}
and the local coordinate transformation, here written as a push-forward,
is 
\begin{equation}
\tilde{V}{}^{i}(\tilde{x})=\sum_{j=1}^{m}\dfrac{\partial\tilde{x}{}^{i}}{\partial x^{j}}V^{j}(x)\label{eq:aa17}
\end{equation}

In mappings between Riemannian manifolds $\phi:(\op M,\br g)\rightarrow(\op N,\br h)$,
the metric tensor also can be equivalently pulled back $\br g=\phi^{*}\tilde{\br g}$
or pushed forward $\tilde{\br g}=\phi_{*}\br g$ . The definition
is 
\begin{equation}
\br g(p)\bigl\{\br V(p),\br W(p)\bigr\}=\tilde{\br g}(\tilde{p})\bigl\{\tilde{\br V}(\tilde{p}),\tilde{\br W}(\tilde{p})\bigr\}\label{eq:aa18}
\end{equation}
for any general pair of tangent vectors. The component relation, here
expressed as a pull-back, is
\begin{equation}
g_{ij}(x)=\sum_{k,l=1}^{m}\tilde{g}{}_{kl}(\tilde{x})\dfrac{\partial\tilde{x}{}^{k}}{\partial x^{i}}\dfrac{\partial\tilde{x}{}^{l}}{\partial x^{j}}\label{eq:aa19}
\end{equation}

The pushed forward metric $\tilde{\br g}=\phi_{*}\br g$ may or may
not be the same as a pre-existing metric $\br h$ of manifold $\op N$.
The case in which $\phi$ is diffeomorphic (as is assumed here), and
also $\tilde{\br g}=\phi_{*}\br g=\br h$, is called an \emph{isometry.} 

As illustrated in \prettyref{fig:Act-Pass-fig}, an active diffeomorphism
is intended to transform the objects representing the physical world,
but keep the reference system unchanged. This requires the target
manifold to be the same as the original one, $\op N=\op M$, and the
system of local coordinates after the mapping to be the same as before,
$\tilde{\psi}=\psi$. The relation between old and new local coordinate
values is defined above as $\tilde{x}=\theta(x)$ where $\theta=\tilde{\psi}\circ\phi\circ\psi^{-1}$.
When $\tilde{\psi}=\psi$, this becomes 
\begin{equation}
\theta=\psi\circ\phi\circ\psi^{-1}\label{eq:aa20}
\end{equation}
which is a defining property of any active diffeomorphism. Note that,
unlike the passive case in \prettyref{subsec:passive} which only
changed the local coordinates while leaving the underlying manifold
objects unchanged, active diffeomorphisms change the underlying manifold
objects in $\op M$ to new underlying manifold objects in the same
manifold $\op M$. 

\subsection{Active Diffeomorphism with Fixed Metric}

In the pre-general-relativistic context of standard differential geometry,
metric $\br g$ is a fixed part of the definition of a Riemannian
manifold, denoted $(\op M,\br g)$, and the metric $\br h$ is a fixed
part of the definition of the target manifold $(\op N,\br h).$ Since
active diffeomorphisms are automorphisms with $\op N=\op M$, and
since a metric is fixed to its manifold, it must also be true that
$\br h=\br g$. Thus the mapping is 
\begin{equation}
\phi:(\op M,\br g)\rightarrow(\op M,\br g)\label{eq:aa21}
\end{equation}
This means that only\emph{ isometric} active diffeomorphisms are allowable
in this pre-general-relativistic context, those with $\tilde{\br g}=\phi_{*}\br g=\br g$.\footnote{Due to the condition $\tilde{\psi}=\psi$ in \prettyref{eq:aa20},
the local coordinate expression of this isometry is $\tilde{g}{}_{ij}(\tilde{x})=g_{ij}(\tilde{x})$.} 

\subsection{Active Diffeomorphism in General Relativity}

In general relativity the metric is not a fixed, prescribed property
of a Riemannian manifold. It is the solution of a differential equation,
unknown until the equation is solved. Thus, as developed by \citet{wald}
and \citet{carroll}, in general relativity the metric tensor can
be transformed arbitrarily in active diffeomorphisms, just as one
would transform any other second rank, covariant tensor field. The
active diffeomorphism can be represented in standard notation as
\begin{equation}
\phi:(\op M,\br g)\rightarrow(\op M,\phi_{*}\br g)\label{eq:aa22}
\end{equation}
If $\phi_{*}\br g=\br g$ we are of course back to the isometric transformations
of \prettyref{eq:aa21}.\footnote{Appendix 5 of \citet{carroll} refers to the mappings in \prettyref{eq:aa22}
simply as \textquotedbl{}diffeomorphisms\textquotedbl{} or, occasionally,
\textquotedbl{}active diffeomorphisms.\textquotedbl{} Thus Carroll's
term \textquotedbl{}diffeomorphism\textquotedbl{} is a synonym of
\textquotedbl{}active diffeomorphism.\textquotedbl{} Also, Wald and
Carroll refer to passive diffeomorphisms as gauge transformations.} Although $\br g$ may transform in the active diffeomorphisms of
general relativity, they remain automorphisms from manifold $\op M$
to itself without change of local coordinate system; thus \prettyref{eq:aa20}
must remain true for them.

\subsection{Generation of Active Diffeomorphisms}

\label{subsec:gen}Some of the machinery of Lie Group theory can be
borrowed to generate active diffeomorphisms from given tangent vector
fields. A useful class of   active diffeomorphisms can be constructed
by considering the mapping $\phi_{\tau}$ along a given tangent vector
field $V(x)$.\footnote{Section 39 of \citet{arnold}, pages 68-70 and Chapter 9 of \citet{lee-smooth},
and pages 27-32 and 250-251 of \citet{oneill}.} Given a chosen starting point $p\in\op M$, a smooth mapping $\eta:(0,\tau{}_{1})\rightarrow\op M$
defines a curve in $\op M$, with $p(\tau)=\eta(\tau)$ and $p(0)=\eta(0)=p$.
In terms of local coordinates this is $x(\tau)=\psi(p(\tau))$. Differentiating
this curve with respect to $\tau$ gives what is sometimes called
a \textquotedbl{}velocity\textquotedbl{} tangent vector $W(x(\tau))$
along the curve. Its components are $\dot{x}^{i}(\tau)=dx^{i}(\tau)/d\tau$.
Given a general tangent vector field $V(x),$ a curve whose velocity
matches that tangent vector for every $\tau\in(0,\tau_{1})$ is defined
by the set of differential equations 
\begin{equation}
\dot{x}^{i}(\tau)=V^{i}(x(\tau))\quad\quad\mbox{where}\quad\quad i=1,\ldots,m\label{eq:aa23}
\end{equation}
whose solution $x(\tau)$ can be described as an integral curve or
\textquotedbl{}field line\textquotedbl{} of $V(x)$ passing through
$x(0)$. The corresponding field line in the manifold is then $p(\tau)=\psi^{-1}(x(\tau))$.

Since the tangent vector field is assumed to be defined at all points
of $\op M$, we can consider the family of all such field-line curves
beginning at every point $p\in M$. Consider an active diffeomorphic
mapping $\phi_{\tau}:\op M\rightarrow\op M$ which simultaneously
carries each $p=p(0)$ in $\op M$ into a $\tilde{p}=p(\tau)$ along
the particular field line starting at $p$. When $\tau=0$, this mapping
is the identity mapping $\phi_{0}=I$. When $\tau>0$, mapping $\phi_{\tau}$
will move each point $p=p(0)$ of $\op M$ along the appropriate field
line to a new point $\tilde{p}=p(\tau)=\phi_{\tau}(p)$. Expressing
the same mapping in local coordinates, each point $x=x(0)=\psi(p)$
is moved by active diffeomorphism $\theta_{\tau}=\psi\circ\phi_{\tau}\circ\psi^{-1}$
into a new point $\tilde{x}=x(\tau)=\theta_{\tau}(x)$. It is important
that the mapping $\phi_{\tau}$ is smoothly connected to the identity
at $\tau=0$. This ensures that the generated active diffeomorphisms
based on $\phi_{\tau}$ do not involve a change of coordinate scheme
that would violate \prettyref{eq:aa20}.

If $V(x)$ is a Killing Vector Field, then, by definition the active
diffeomorphism $\phi_{\tau}$ is isometric. Generation of more general
active diffeomorphisms with $\tilde{\br g}=\phi_{*}\br g\neq\br g$
requires that $V(x)$ not be a Killing Vector Field.

\subsection{Examples of the Generation of Active Diffeomorphisms}

Consider Cartesian three space with coordinates\footnote{The coordinates $x=(x^{1},x^{2},x^{3})$ are written here as $(x,y,z)$
for readibility.} $(x,y,z)$ and metric defined by the matrix $\mat g(x)=\text{diag}(1,1,1)$. 

Choose a Killing Vector Field with components $V(x)=(-y,x,0)$. Then
\prettyref{eq:aa23} becomes 
\begin{equation}
\dfrac{dx(\tau)}{d\tau}=-y\quad\quad\quad\dfrac{dy(\tau)}{d\tau}=x\quad\quad\quad\dfrac{dz(\tau)}{d\tau}=0\label{eq:aa24}
\end{equation}
with solution
\begin{equation}
\tilde{x}=x(\tau)=A\cos\tau-B\sin\tau\quad\quad\tilde{y}=y(\tau)=A\sin\tau+B\cos\tau\quad\quad\tilde{z}=C\label{eq:aa25}
\end{equation}

The initial condition $(x(0),y(0),z(0))=(x,y,z)$ then gives the active
diffeomorphism for epoch $\tau$ as
\begin{equation}
\tilde{x}=x\cos\tau-y\sin\tau\quad\quad\tilde{y}=x\sin\tau+y\cos\tau\quad\quad\tilde{z}=z\label{eq:aa26}
\end{equation}
which is the same as \prettyref{eq:a3} with epoch $\tau$ identified
with rotation angle $\alpha$. Since this transformation is orthogonal,
the transformed metric remains $\tilde{\mat g}(\tilde{x})=\text{diag}(1,1,1)$.
Thus $\tilde{\br g}=\phi_{*}\br g=\br g$ and the active diffeomorphism
is isometric.

Now choose a non-Killing Vector Field with components $V(x)=(y,0,0)$.
Then \prettyref{eq:aa23} becomes 
\begin{equation}
\dfrac{dx(\tau)}{d\tau}=y\quad\quad\quad\dfrac{dy(\tau)}{d\tau}=0\quad\quad\quad\dfrac{dz(\tau)}{d\tau}=0\label{eq:aa27}
\end{equation}
with solution
\begin{equation}
\tilde{x}=x(\tau)=a+b\tau\quad\quad\tilde{y}=y(\tau)=b\quad\quad\tilde{z}=c\label{eq:aa28}
\end{equation}
The initial condition $(x(0),y(0),z(0))=(x,y,z)$ then gives the active
diffeomorphism for epoch $\tau$ as
\begin{equation}
\tilde{x}=x+y\tau\quad\quad\tilde{y}=y\quad\quad\tilde{z}=z\label{eq:aa29}
\end{equation}
The transformed metric obtained from the inverse of \prettyref{eq:aa19}
is 
\begin{equation}
\tilde{\mat g}(\tilde{x})=\left(\begin{array}{ccc}
1 & -\tau & 0\\
-\tau & (\tau^{2}+1) & 0\\
0 & 0 & 1
\end{array}\right)\label{eq:aa30}
\end{equation}
and the active diffeomorphism is not isometric.

Note that both of these active diffeomorphisms reduce smoothly to
the identity when $\tau=0$, consistent with \prettyref{eq:aa20}
and the condition that active diffeomorphisms do not change the system
of local coordinates but only the manifold point being represented.

\subsection{Essential Difference Between Passive and Active Diffeomorphisms}

\label{subsec:padiff}Passive diffeomorphisms change the system of
local coordinates but do not change the manifold objects being represented
in those coordinates. Thus the same point $p\in\op M$ in the manifold
is represented by $x=\psi(p)$ and $x'=\psi'(p)$. Also $g_{\mu\nu}(x)$
and $g'_{\mu\nu}(x')$ both represent the same underlying metric tensor
field $\br g(p)$ defined on the manifold $\op M$.

Active diffeomorphisms are the opposite of passive ones. In them the
underlying manifold point $p\in\op M$ is mapped to a different point
$\tilde{p}=\phi_{\tau}(p)\in\op M$ with both old and new points being
represented in the same system of local coordinates. Thus $x=\psi(p)$
and $\tilde{x}=\psi(\tilde{p})$ are different, not because of a change
of coordinate system, but because the manifold point being represented
has been mapped from $p$ to $\tilde{p}$. The underlying manifold
objects are also changed: $\tilde{f}(p)\neq f(p)$, $\tilde{\br V}(p)\neq\br V(p)$,
$\tilde{\br g}(p)\neq\br g(p)$, and so on for other tensors. Manifold
objects are often used to model the physical world. The change of
these manifold objects by an active diffeomorphism thus changes the
model. The thrust of the hole argument in \prettyref{sec:hole-in-gr}
is to prove that two different models, one derived from the other
by an active diffeomorphism, can both be solutions of the same Einstein
field equation.

\section{Einstein's Hole Argument in General Relativity}

\label{sec:hole-in-gr}The Einstein field equation may be written
as
\begin{equation}
R_{\mu\nu}-\dfrac{1}{2}g_{\mu\nu}R+\kappa T_{\mu\nu}=0\label{eq:gr1}
\end{equation}
where $R_{\mu\nu}$ is the Ricci tensor, $R$ is the curvature invariant,
$\kappa$ is a universal constant related to the Newton gravitational
constant, and $T_{\mu\nu}$ is the energy-momentum tensor source.
This equation may be written in a form that presents its dependency
on the metric tensor and its derivatives explicitly. It is 
\begin{equation}
\op R_{\mu\nu}(g(x),x)-\dfrac{1}{2}g_{\mu\nu}(x)\left(g^{\alpha\beta}(x)\op R_{\alpha\beta}(g(x),x)\right)+\kappa T_{\mu\nu}(x)=0\label{eq:gr2}
\end{equation}
where the functions $\op R_{\mu\nu}$ are defined by\footnote{See \citet{synge} equations 2.241, 2.242, and 3.203. The Einstein
summation convention is used. A term containing a repeated Greek index
is summed over that index, from $0$ to $3$. }
\begin{equation}
\op R_{\mu\nu}(g(x),x)=\dfrac{\partial}{\partial x^{\nu}}\left\{ {\alpha\atop \mu\alpha}\right\} -\dfrac{\partial}{\partial x^{\alpha}}\left\{ {\alpha\atop \mu\nu}\right\} +\left\{ {\beta\atop \mu\alpha}\right\} \left\{ {\alpha\atop \beta\nu}\right\} -\left\{ {\beta\atop \mu\nu}\right\} \left\{ {\alpha\atop \beta\alpha}\right\} \label{eq:gr3}
\end{equation}
\begin{equation}
\text{where}\quad\quad\left\{ {\mu\atop \nu\alpha}\right\} =\dfrac{1}{2}g^{\mu\delta}(x)\left(\frac{\partial g_{\delta\nu}(x)}{\partial x^{\alpha}}+\frac{\partial g_{\delta\alpha}(x)}{\partial x^{\nu}}-\frac{\partial g_{\nu\alpha}(x)}{\partial x^{\delta}}\right)\label{eq:gr4}
\end{equation}

Now perform a general non-Killing active diffeomorphism $\theta_{\tau}=\psi\circ\phi_{\tau}\circ\psi^{-1}$
from local coordinates $x$ to local coordinates $\tilde{x}=\theta_{\tau}(x)$.
After this active diffeomorphism the field equation becomes 
\begin{equation}
\op R_{\mu\nu}(\tilde{g}(\tilde{x}),\tilde{x})-\dfrac{1}{2}\tilde{g}_{\mu\nu}(\tilde{x})\left(\tilde{g}^{\alpha\beta}(\tilde{x})\op R_{\alpha\beta}(\tilde{g}(\tilde{x}),\tilde{x})\right)+\kappa\tilde{T}_{\mu\nu}(\tilde{x})=0\label{eq:gr5}
\end{equation}
Comparing \prettyref{eq:gr5} after the active diffeomorphism to \prettyref{eq:gr2}
before it, note that there is no tilde on the function $\op R$. Due
to \prettyref{eq:aa22} and the general covariance of the Ricci tensor,
$\op R_{\mu\nu}(\tilde{g}(\tilde{x}),\tilde{x})$ is exactly the same
function of $(\tilde{g}(\tilde{x}),\tilde{x})$ as $\op R_{\mu\nu}(g(x),x)$
is of $(g(x),x)$. But, since we have not yet applied Einstein's restricted
definition of the energy-momentum source, $\tilde{T}_{\mu\nu}$ is
generally a different function from $T_{\mu\nu}$.

\begin{figure}[H]
\includegraphics[width=1\textwidth]{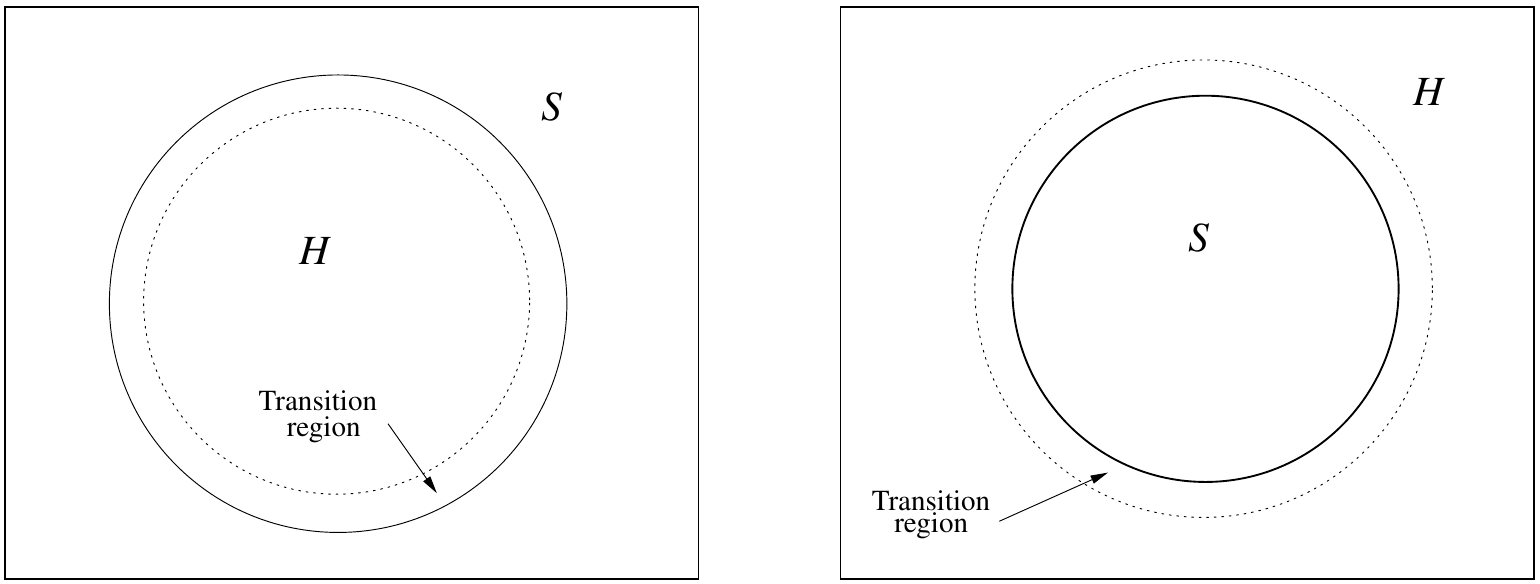}

\caption{\label{fig:hs}A schematic diagram of the hole $H$ and source $S$
regions of the hole argument. On the left the hole $H$ is surrounded
by source $S$ as envisioned by Einstein. On the right, the source
$S$ is surrounded by the hole $H$ as used in \prettyref{sec:schole}.
The transition regions inside $H$ are necessary to preserve the differentiability
of the active diffeomorphism.}
\end{figure}
Now apply Einstein's restrictions. First, assume only experimental
situations in which there is a hole region $H$ with $T_{\mu\nu}(x)\equiv0$
for $x\in H$. Then consider an active diffeomorphism that is the
identity ($V(x)\equiv0$ and hence $\theta_{\tau}(x)=I$) in the region
$S$ that is the complement of $H$, but not the identity in $H$
itself.\footnote{See \prettyref{fig:hs}. The chosen active diffeomorphism must have
a small transition region just inside $H$, transitioning smoothly
from identity in $S$ to non-identity inside $H$, in order to satisfy
the basic condition that active diffeomorphisms must be smooth functions. } It follows that the energy-momentum tensor is untransformed both
in $H$ (because a zero tensor transforms to the zero tensor regardless
of the transformation) and $S$ (because the active diffeomorphism
is the identity in $S$). Thus $\tilde{T}_{\mu\nu}(\tilde{x})=T_{\mu\nu}(\tilde{x})$
throughout the manifold, and \prettyref{eq:gr5} becomes
\begin{equation}
\op R_{\mu\nu}(\tilde{g}(\tilde{x}),\tilde{x})-\dfrac{1}{2}\tilde{g}_{\mu\nu}(\tilde{x})\left(\tilde{g}^{\alpha\beta}(\tilde{x})\op R_{\alpha\beta}(\tilde{g}(\tilde{x}),\tilde{x})\right)+\kappa T_{\mu\nu}(\tilde{x})=0\label{eq:gr6}
\end{equation}
with no tilde on the $T$. However, the function $\tilde{g}_{\mu\nu}$
that solves \prettyref{eq:gr6} is not the same function as the $g_{\mu\nu}$
that solves \prettyref{eq:gr2}. Inside the hole region $\tilde{g}_{\mu\nu}(\tilde{x})\neq g_{\mu\nu}(\tilde{x})$.

If \prettyref{eq:gr6} is satisfied, it follows that the differential
equation
\begin{equation}
\op R_{\mu\nu}(\tilde{g}(x),x)-\dfrac{1}{2}\tilde{g}_{\mu\nu}(x)\left(\tilde{g}^{\alpha\beta}(x)\op R_{\alpha\beta}(\tilde{g}(x),x)\right)+\kappa T_{\mu\nu}(x)=0\label{eq:gr7}
\end{equation}
must also be satisfied. Comparison of \prettyref{eq:gr7} and \prettyref{eq:gr2}
demonstrates that $\tilde{g}_{\mu\nu}(x)$ and $g_{\mu\nu}(x)$ are
both solutions to the same Einstein field equation. Thus, in this
experimental situation, there are two or more solutions to the Einstein
field equation with the same energy-momentum tensor source, as Einstein
asserted. The physical meaning of these multiple solutions, and the
possibility of the rejection of some of them as spurious, is the subject
of \prettyref{sec:phys} below.

Note that \prettyref{eq:gr7} differs from \prettyref{eq:gr6} only
in the replacement $\tilde{x}\rightarrow x$ throughout. The argument
leading from \prettyref{eq:gr6} to \prettyref{eq:gr7} is as follows:
In linear algebra, one often uses dummy indices whose replacement
by other letters does not change a sum, provided that the two sets
of indices are summed over the same range. Thus the equality $\sum_{i=1}^{5}K_{i}=6$
is true if and only if $\sum_{j=1}^{5}K_{j}=6$ is true. Dummy indices
have an analog in differential equations. The equality $df(t)/dt=-\lambda f(t)$
is true if and only if $df(u)/du=-\lambda f(u)$ is true, provided
only that the dummy variables $t$ and $u$ are of the same character
and range, here real numbers in $(-\infty,\infty)$. Now make the
same sort of substitution in \prettyref{eq:gr6}, with $\tilde{x}$
in place of $t$ and $x$ in place of $u$, with $\tilde{g}_{\mu\nu}$
playing the role of $f$. The condition that $\tilde{x}$ and $x$
are variables of the same character and range is ensured by the condition
$\tilde{\psi}=\psi$ and \prettyref{eq:aa20}. Also, the fact that
active diffeomorphisms constructed as in \prettyref{subsec:gen} are
smoothly connected to the identity when $\tau\rightarrow0$ rules
out transformations, such as from Cartesian to spherical polar, that
would make the ranges of $\tilde{x}$ and $x$ different. With the
substitution $\tilde{x}\rightarrow x$, the equality in \prettyref{eq:gr6}
is true if and only if the equality in \prettyref{eq:gr7} is true.

Note the crucial importance of Einstein's restriction that the energy-momentum
source must vanish in the hole. Without that restriction, the $T$
in both \prettyref{eq:gr6} and \prettyref{eq:gr7} would be replaced
by $\tilde{T}$. The $\tilde{g}$ would still be a different metric
solution, but it would be the solution to a \emph{different} differential
equation, one with an actively transformed source $\tilde{T}$ that
models a \emph{different} experimental situation, and not a second
solution to the original differential equation with the original source
$T$. Without the Einstein condition on $T$, the above proof of multiple
solutions fails.

\section{Physical Meaning\protect\footnote{As noted above, by \textquotedbl{}physical meaning\textquotedbl{}
of local coordinates I mean a defined relation between them and some
physical quantity like length or relativistic interval.} in Einstein's Multiple Metrics}

\label{sec:phys}Einstein's final form of his field equation is generally
covariant. It therefore suffers from the multiplicity of  solutions
derived in \prettyref{sec:hole-in-gr}. His resolution was to assert
that all the metric solutions are physically equivalent, and to deny
that local coordinates represent anything real.\footnote{\label{fn:stachel}See \textquotedbl{}How Einstein Discovered General
Relativity: A Historical Tale With Some Contemporary Morals\textquotedbl{}
pp. 293-299 of \citet{stachel-btoz}}

In one reading, Einstein's denial of the reality of local coordinates
only repeats a fact of pre-general-relativistic differential geometry.
The local coordinates $x=\psi(p)$ with $p\in\op M$ defined on a
bare manifold $\op M$ by means of homeomorphism $\psi$ do not initially
have any definite relation to any physical or geometrical quantity.
The coordinates $x$ are just $m$-tuples of real numbers. In pre-general-relativistic
cases, these numbers acquire geometric or physical meaning only when
the move is made to Riemannian geometry by adding a (fixed) metric
to the manifold.\footnote{Of course the natural Euclidean metric of $m$-tuples of real numbers
is always available. But it does not have to be applied. For example,
in Hamiltonian mechanics the Euclidean metric is not used, with a
symplectic structure function applied instead.} The situation is even more extreme in general relativity, in which
a definite metric is not even available to be applied to $\op M$
until \emph{after} the field equation is solved. In general relativity,
a solution to the Einstein field equation is to be obtained using
local coordinates of unknown physical meaning. Each metric tensor
solution then has a privileged role; each of them determines the physical
meaning of the coordinates $x$ in terms of which it is written.

Since local coordinates obtain their physical meaning only from a
metric solution to the field equation, it follows that different metric
solutions to the field equations may give different physical meanings
to the same set of local coordinates. The fact that, at a given manifold
point $p$, the local coordinates $x=(x^{0},x^{1},x^{2},x^{3})$ in
$g_{\mu\nu}(x)$ are the same quadruple of real numbers as the $x$
in $\tilde{g}_{\mu\nu}(x)$ does not mean that the local coordinates
$x$ have the same physical meaning in both solutions. Each metric
solution brings its own assignment of physical or geometrical meaning
to the local coordinates used to write it. 

On this reading, Einstein's statement should be modified to say not
that local coordinates have no meaning,\footnote{Note that solutions to the Einstein field equation such as the Schwarzschild
or Robertson-Walker metrics do indeed assign a physical meaning to
their coordinates.} but rather that local coordinates have no \emph{independent} meaning,
independent of the metric solution. Each of the multiple metric solutions
carries its own physical interpretation of its own local coordinates.
I propose three resolutions to this problem of undetermined local
coordinate meaning, each of which denies the necessity of multiple
solutions to the field equation.

\subsection{Resolution A: Active Diffeomorphisms Must be Isometric}

Resolution A suggests that a strict definition of the term \textquotedbl{}active
diffeomorphisms\textquotedbl{} requires them to be isometric, and
thus prevents their use in the hole argument. The condition $\tilde{\psi}=\psi$
leading to \prettyref{eq:aa20} was to guarantee that the mapping
$\psi(p)$ from manifold points $p$ to local coordinates is the same
before and after the active diffeomorphism. This is the defining condition
that an active diffeomorphism modifies the physical world but must
not modify the system of coordinates used to observe it. But when
the non-isometric case $\tilde{\br g}\neq\br g$ is allowed, the transformed
metric $\tilde{\br g}(\tilde{x})$ gives transformed coordinate $\tilde{x}$
a physical meaning different from the one that the original metric
$\br g(x)$ gave to original coordinate $x$. This difference of meaning
modifies the system of local coordinates in an essential way; it therefore
violates the defining condition of active diffeomorphisms and must
be rejected. But when only isometric active diffeomorphisms are allowed,
there is no hole argument. By definition, an isometric active diffeomorphism
simply replicates the same metric tensor and does not provide a new
one; multiple solutions are not generated.

\subsection{Resolution B: Selection by Symmetry}

Resolution B ignores the strict definition demanded by Resolution
A and allows non-isometric active diffeomorphisms of the sort outlined
in \prettyref{sec:actpas} and \citet{carroll}. Each of the resulting
multiple solutions to \prettyref{eq:gr2} is then an equally valid
candidate solution, but each gives a different physical meaning to
the local coordinates in terms of which it is written. As with many
differential equations, one must then reject some solutions as spurious.
Any solution that gives a physical meaning to its local coordinates
that violates the symmetries demanded by the experimental situation
being modeled can be rejected as spurious. Thus solving \prettyref{eq:gr2}
is only the first step in a solution procedure for the Einstein field
equation. There are multiple solutions, but also a method to select
the correct one from among them and to reject the others as spurious.

\subsection{Resolution C: Use of a Template}

Resolution C is similar to Resolution B above, but rather than actually
choosing one solution with the desired symmetry from a multiplicity
of candidate solutions, one simply enforces symmetry from the start
by specifying a template that forces a single solution exhibiting
that symmetry. For example, in \prettyref{sec:schwarz} for the Schwarz\-schild
metric one solves the Einstein equation in two steps, the first of
which is to choose a template metric which forces spherical symmetry
and almost completely defines the physical meanings of its coordinates.
The second step is to substitute this template into the field equation
to determine its remaining parameters. In effect, the metric is largely
determined by the first step; the Einstein field equation is used
as a kind of auxiliary condition to determine certain residual parameters
and ensure consistency with Newtonian gravity. Solutions other than
the one derived from the template then violate the template and its
symmetry and can be rejected as spurious. 

Although Resolutions A, B, and C differ, they agree that, despite
the mathematical proof in \prettyref{sec:hole-in-gr}, the existence
of a unique solution to the Einstein field equation cannot be ruled
out.

\section{The Schwarzschild Example}

\label{sec:schwarz}A good test case to illustrate the hole argument
with active diffeomorphisms that modify the metric tensor is the Schwarzschild
solution to the Einstein field equation in the empty space surrounding
a spherically symmetric source region.

The first step to the Schwarzschild solution is to construct a template
metric that defines the geometric properties of some of the local
coordinates and enforces spherical symmetry.\footnote{See Chapter 8 of \citet{weinberg}, Chapter 11 of \citet{rindler},
Chapter 14 of \citet{dinverno}.} A standard template denotes the variable set as $(x^{0},x^{1},x^{2},x^{3})=(t,r,\theta,\phi)$
and sets the template $g_{\mu\nu}(x)$ equal to the diagonal matrix
\begin{equation}
\mat g(x)=\text{diag}\left(-c^{2}\beta(r),\alpha(r),r^{2},r^{2}\sin^{2}\theta\right)\label{eq:sch1}
\end{equation}
This choice of template enforces the spherical symmetry of the problem,
identifies $\theta$ and $\phi$ as the standard angles of spatial
spherical polar coordinates, and makes the area of the surface $t=\text{const.}$,
$r=\text{const.}$ equal to $4\pi r^{2}$. This template is substituted
into \prettyref{eq:gr2}; straightforward algebra then determines
the functions $\alpha$ and $\beta$ and arrives at
\begin{equation}
\mat g(x)=\mbox{diag}\left\{ -\left(1-\dfrac{2m}{r}\right)c^{2},\left(1-\dfrac{2m}{r}\right)^{-1},r^{2},r^{2}\sin^{2}\theta\right\} \label{eq:sch2}
\end{equation}
where $m=GM/c^{2}$, with Newton's gravitational constant $G$, the
total mass of the source $M$, and the speed of light $c$.\footnote{If the unknown functions $\alpha$ and $\beta$ in \prettyref{eq:sch1}
are allowed to be functions of both $r$ \emph{and} $t$, the same
metric \prettyref{eq:sch2} is obtained, a result known as Birkhoff's
theorem. See Chapter XV of \citet{birkhoff}, and Chapter 14 of \citet{dinverno}.} The Schwarzschild solution in \prettyref{eq:sch2} is \emph{uniquely
determined} given the template that sets its desired symmetry. 

The Robertson-Walker metric is similarly derived from a template enforcing
its symmetries\footnote{Chapter 13 of \citet{weinberg} derives this template as well as \prettyref{eq:sch1}
from a requirement of maximal subspace symmetry, with no prior reference
to the Einstein field equation.} 
\begin{equation}
\mat g(x)=\text{diag}\left(-c^{2},(\alpha(t))^{2}/(1-kr^{2}),(r\alpha(t))^{2},(r\alpha(t))^{2}\sin^{2}\theta\right)\label{eq:sch3}
\end{equation}
where $k$ is $-1$, $0$, or $+1$ and scale factor $\alpha(t)$
can be derived from the Einstein field equation together with assumptions
about the density and nature of matter in a cosmological model.\footnote{See \citet{dinverno}, Section 22.9.}

\section{The Hole Argument with the Schwarzschild Solution}

\label{sec:schole}Now apply the hole argument to the Schwarz\-schild
solution. Referring to \prettyref{fig:hs} and the description of
the hole argument in \prettyref{sec:hole-in-gr}, region $S$ can
be taken as all points with $r\le r_{1}$ where $r_{1}$ is a radius
beyond all energy-momentum tensor sources and also beyond the Schwarzschild
radius $r=2m$. Region $H$ is then all points with $r>r_{1}$. The
transition region inside $H$ is all points with $r_{1}<r<r_{2}$
where $r_{2}$ is some arbitrarily chosen boundary. In this example,
the \textquotedbl{}hole\textquotedbl{} region $H$ is in fact exterior
to the source region $S$, but this choice makes no difference to
the hole argument. All that is required is that $H\cap S=\textrm{Ø}$
and $H\cup S=\op M$.

To apply the hole argument, first define a smoothing function to enforce
the differentiability of the active diffeomorphism in the transition
region. It is\footnote{See pages 40-42 of \citet{lee-smooth}.} 
\begin{equation}
\xi(s)=\left\{ \begin{array}{ccc}
\exp(-1/s) & \mbox{for} & s>0\\
\: & \: & \:\\
0 & \mbox{for} & s\le0
\end{array}\right.\label{eq:schole1}
\end{equation}
Then choose an arbitrary (but non-Killing) tangent vector field $X(x)$
and define a tangent vector field $V(x)$ by
\begin{equation}
V(x)\equiv0\quad\quad V(x)=\left(\dfrac{\xi(r-r_{1})}{\xi(r-r_{1})+\xi(r_{2}-r)}\right)X\left(x\right)\quad\quad V(x)=X(x)\label{eq:schole2}
\end{equation}
in region $S$, the transition region, and the remainder of region
$H$, respectively.

As described in \prettyref{subsec:gen}, for any fixed $\tau$ value
an   active diffeomorphism $\phi_{\tau}$ can be defined by following
the field lines of tangent vector $V(x)$. It changes the metric of
\prettyref{eq:sch2} to a new metric in region $H$, but without changing
the metric or the source in region $S$ where $V(x)\equiv0$ and hence
$\phi_{\tau}=I,$ the identity transformation. 

Applying the hole argument with this   active diffeomorphism, the
unchanged mass source in region $S$ now produces a family of different
metrics $\tilde{\br g}=\phi_{\tau*}\br g$ that solve \prettyref{eq:gr2}
in region $H$. It then follows that a given source in region $S$
of the Schwarzschild problem produces many different solutions in
region $H$, one for each $\tau$ value and choice of tangent vector
field $X(x)$. 

For example, choose the $X(x)$ to have local coordinates $(0,a\phi,0,0)$
where $a$ is some fixed parameter having units of length. Use the
inverse of \prettyref{eq:aa19} with the  active diffeomorphism given
by the procedure in \prettyref{subsec:gen} to write the transformed
metric tensor. In region $H$ beyond the transition region, it is
\begin{equation}
\tilde{\mat g}(x)=\left(\begin{array}{cccc}
-c^{2}\lambda & 0 & 0 & 0\\
0 & \lambda^{-1} & 0 & -a\tau\lambda^{-1}\\
0 & 0 & \left(r-a\tau\phi\right)^{2} & 0\\
0 & -a\tau\lambda^{-1} & 0 & \chi
\end{array}\right)\label{eq:schole3}
\end{equation}
where
\begin{equation}
\lambda=\dfrac{r-a\tau\phi-2m}{r-a\tau\phi}\quad\quad\text{and}\quad\quad\chi=a^{2}\tau^{2}\lambda^{-1}+\left(r-a\tau\phi\right)^{2}\sin^{2}\theta\label{eq:schole4}
\end{equation}

Because of its generation by the hole procedure, \prettyref{eq:schole3}
is certainly another  solution to the Einstein field equation with
the same source field $T$. But it may be rejected by symmetry considerations.
The Schwarzschild solution \prettyref{eq:sch2} enforces the desired
spherical symmetry resulting from the assumed spherically symmetric
mass distribution. The metric solution \prettyref{eq:schole3} lacks
that spherical symmetry. In fact, due to the uniqueness of the Schwarzschild
solution given the template metric, there is no possible alternate
solution with the same template \prettyref{eq:sch1} but different
parameters $\alpha$ and $\beta$. The Schwarzschild metric stands
as a counterexample to the proposition that the Einstein field equation
must of necessity always have multiple solutions.

The Robertson-Walker metric is similarly derived by starting with
a template, \prettyref{eq:sch3}, enforcing spherical symmetry. A
non-isometric active diffeomorphism applied to it will result in a
metric that violates that template, just as in the Schwarzschild case. 

\section{Conclusion}

\label{sec:con}The proof in \prettyref{sec:hole-in-gr} demonstrates
mathematically that the Einstein field equation has multiple metric
solutions. But since no metric is defined until after the field equation
is solved, that proof is of necessity just a numerical exercise written
using local coordinates that are quadruples of real numbers with no
definite physical meaning, i.e., no assigned relation to relativistic
interval. After the field equation is solved, each of the multiple
metric solutions then assigns its own physical meaning to the local
coordinates in terms of which it is written. As noted in Resolutions
B and C of \prettyref{sec:phys}, these various physical meanings
of the local coordinates, as read from the various metric solutions,
may then be used to reject as spurious those solutions whose local
coordinates have a meaning inconsistent with the symmetries of the
experimental situation being modeled. This rejection of spurious solutions
opens the possibility that in some cases, such as the Schwarzschild
metric, only one solution may survive. Thus the hole argument cannot
prove the assertion that the Einstein field equation must have multiple
solutions.

A considerable intellectual superstructure has been built on the foundation
of the hole argument, beginning with Einstein himself who asserted
that because of it the local coordinates used to write his field equation
can have no physical meaning.\footnote{See \prettyref{sec:phys}.}
Later authors\footnote{\citet{earman-norton}, Chapter V of \citet{stachel-btoz} and others
quoted therein.} have expanded this intuition into a general argument against what
is sometimes called manifold substantivalism, roughly defined as a
realist interpretation of the manifold of differential geometry. The
failure of the hole proof noted above removes Einstein's contribution
to this argument.

\bibliographystyle{spbasic}
\bibliography{ODJ}

\begin{thebibliography}{24}
\providecommand{\natexlab}[1]{#1}
\providecommand{\url}[1]{{#1}}
\providecommand{\urlprefix}{URL }
\expandafter\ifx\csname urlstyle\endcsname\relax
  \providecommand{\doi}[1]{DOI~\discretionary{}{}{}#1}\else
  \providecommand{\doi}{DOI~\discretionary{}{}{}\begingroup
  \urlstyle{rm}\Url}\fi
\providecommand{\eprint}[2][]{\url{#2}}

\bibitem[{Arnold(1978)}]{arnold}
Arnold VI (1978) Mathematical Methods of Classical Mechanics. Springer, New
  York

\bibitem[{Birkhoff and Langer(1923)}]{birkhoff}
Birkhoff GD, Langer RE (1923) Relativity and Modern Physics. Harvard University
  Press, Harvard, MA

\bibitem[{Carroll(2016)}]{carroll}
Carroll SM (2016) Spacetime and Geometry. Pearson Education Ltd., London

\bibitem[{d'Inverno(1992)}]{dinverno}
d'Inverno R (1992) Introducing Einstein's Relativity. Oxford University Press,
  Oxford, UK

\bibitem[{Earman and Norton(1987)}]{earman-norton}
Earman J, Norton J (1987) What price substantivalism? {T}he hole story. Brit J
  Phil Sci 38:515--525

\bibitem[{Einstein and Grossmann(1913)}]{EinsteinGrossmann}
Einstein A, Grossmann M (1913) Entwurf einer verallgemeinerten
  {R}elativit{\"a}tstheorie und einer {T}heorie der {G}ravitation. Zeitschrift
  fur Mathematik und Physik 62:225--261

\bibitem[{Frankel(2004)}]{frankel}
Frankel T (2004) The Geometry of Physics, 2nd edn. Cambridge University Press,
  Cambridge, UK

\bibitem[{Iftime and Stachel(2006)}]{IftimeStachel}
Iftime M, Stachel J (2006) The hole argument for covariant theories. General
  Relativity and Gravitation 38:1241--1252

\bibitem[{Johns(2011)}]{oj}
Johns OD (2011) Analytical Mechanics, 2nd edn. Oxford University Press, Oxford,
  UK

\bibitem[{Lee(1997)}]{lee-riemann}
Lee JM (1997) Riemannian Manifolds. Springer, New York

\bibitem[{Lee(2010)}]{lee-top}
Lee JM (2010) Introduction to Topological Manifolds, 2nd edn. Springer

\bibitem[{Lee(2013)}]{lee-smooth}
Lee JM (2013) Introduction to Smooth Manifolds, 2nd edn. Springer, New York

\bibitem[{Norton(2011)}]{norton-encyc}
Norton J (2011) The hole argument. In: Zalta EN (ed) Stanford Encyclopedia of
  Philosophy; Available at <http://plato.stanford.edu/archives/
  fall2011/entries/spacetime-holearg/>

\bibitem[{O'Neill(1983)}]{oneill}
O'Neill B (1983) Semi-Riemannian Geometry. Academic Press, New York

\bibitem[{Rindler(2006)}]{rindler}
Rindler W (2006) Relativity Special, General, and Cosmological, 2nd edn. Oxford
  University Press, Oxford, UK

\bibitem[{Schulman(2016)}]{schulman-homotopy}
Schulman M (2016) Homotopy type theory: A synthetic approach to higher
  equalities. \urlprefix\url{<http://arxiv.org/abs/1601.05035v3>}

\bibitem[{Stachel(1986)}]{Stachel-active}
Stachel J (1986) What a physicist can learn from the history of {E}instein's
  discovery of general relativity. In: Ruffini R (ed) Proceedings of the Fourth
  Marcel Grossmann Meeting on General Relativity, Elsevier, Amsterdam, pp
  1857--1862

\bibitem[{Stachel(2002)}]{stachel-btoz}
Stachel J (2002) Einstein from 'B' to 'Z'. Birkh{\"a}user, Boston, MA

\bibitem[{Synge and Schild(1978)}]{synge}
Synge JL, Schild A (1978) Tensor Calculus. Dover, New York

\bibitem[{Taubes(2011)}]{Taubes}
Taubes CH (2011) Differential Geometry, Bundles, Connections, Metrics and
  Curvature. Oxford University Press, Oxford, UK

\bibitem[{Torretti(1996)}]{torretti}
Torretti R (1996) Relativity and Geometry. Dover, New York

\bibitem[{Wald(1984)}]{wald}
Wald RM (1984) General Relativity. Univ. of Chicago Press, Chicago, IL

\bibitem[{Weatherall(2018)}]{weatherall}
Weatherall JO (2018) Regarding the `hole' argument. Brit J Phil Sci 69:329--350

\bibitem[{Weinberg(1972)}]{weinberg}
Weinberg S (1972) Gravitation and Cosmology. John Wiley and Sons, New York

\end{thebibliography}

\end{document}